\documentstyle[multicol,prl,aps,amsfonts,bm]{revtex}

\newcommand{\ket}[1]{\mbox{$|{#1}\rangle$}}

\def\beq{\begin{equation}}
\def\eeq{\end{equation}}
\def\beqa{\begin{eqnarray}}
\def\eeqa{\end{eqnarray}}

\begin{document}

\title{
Bounds on the generalized entanglement of formation for multi-party systems
\thanks{Supported by the National Natural Science Foundation of China under Grant No. 69773052}}
\author{An Min WANG$^{1,2}$}
\address{Laboratory of Quantum Communication and Quantum Computing\\
University of Science and Technology of China$^1$\\
Department of Modern Physics, University of Science and Technology of China\\
P.O. Box 4, Hefei 230027, People's Republic of China$^2$}

\maketitle

\begin{abstract}
We present a general method to find the upper and lower bounds on the generalized entanglement of formation for multi-party systems. The upper and lower bounds can be expressed in terms of the bi-partite entanglements of formation and/or entropies of various subsystems. The examples for tri- and four-party systems in the both pure states and mixed states are given. We also suggest a little modified definition of generalized entanglement of formation for multi-party systems if EPR pairs are thought of belonging to the set of maximally entangled states. 

\medskip
{\noindent}PACS: 03.67.-a, 03.65.Bz, 89.70.+c\vfill
\end{abstract}

\smallskip

As a resource, entanglement can be exploited to implement novel quantum information processing task \cite{Bennett1,Plenio1,Shor}. So there has been an ongoing effort to characterize quantitatively and qualitatively entanglement \cite{Bennett2,Vedral1,Wootters}, in special, for multi-particles \cite{Cirac} and multi-party system \cite{Plenio2} in recent years. The central issue is now what the definition of maximally entangled states (set) for multi-party systems would be so that one can clearly know to distill to them. Since for bi-partite systems the relative entanglement entropy is a upper bound to distillable entanglement, it is indeed reasonable to think this to be the case for multi-party systems \cite{Plenio2}. In their paper, Plenio and Vedral showed this institution by deriving various upper and lower bounds of the relative entanglement entropy for tri-party systems in terms of the bi-party entanglements of formation and distillation and entropies of various subsystems. In this paper, we will derive out the upper and lower bounds on the generalized entanglement of formation for both tri-party systems \cite{My1} and four-party systems not only in the pure states but also in the mixed states which can be expressed in terms of the bi-party entanglements of formation and entropies of various subsystems. Moreover, our method, in principle, can be extended to find the bounds on the generalized entanglement of formation for the arbitrary multi-party systems. 

We start with the following inequantions for a tri-party quantum system $X,Y,Z$
\beqa
S(\rho_X)+S(\rho_Y)\leq S(\rho_{X\!Z})+S(\rho_{Y\!Z})\label{S31}\\
S(\rho_{X\!Y\!Z})+S(\rho_Y)\leq S(\rho_{X\!Y})+S(\rho_{Y\!Z})\label{S32}
\eeqa
Their proofs are based deep mathematical results known as Lieb's theorem \cite{Lieb}. It is easy to see that
\beqa
S(\rho_{X})+S(\rho_{Y})+S({Z})
&\leq& S(\rho_{X\!Y})+S(\rho_{X\!Z})+S(\rho_{Y\!Z})\\
3S(\rho_{X\!Y\!Z})+S(\rho_{X})+S(\rho_{Y})+S(\rho_{Z})&\leq& 2\left[S(\rho_{X\!Y})+S(\rho_{X\!Z})+S(\rho_{Y\!Z})\right]
\eeqa
For a four-party system we have the corresponding generalization:
\beqa
S(\rho_{A})+S(\rho_{B})+S(\rho_{C})+S(\rho_D)&\leq& \frac{2}{3}[S(\rho_{A\!B})+S(\rho_{A\!C})+S(\rho_{A\!D})+S(\rho_{B\!C})+S(\rho_{B\!D})+S(\rho_{C\!D})]\label{S45}\\
S(\rho_{A})+S(\rho_{B})+S(\rho_{C})+S(\rho_{D})&\leq& S(\rho_{A\!B\!C})+S(\rho_{A\!B\!D})+S(\rho_{A\!C\!D})+S(\rho_{B\!C\!D})\label{S46}
\eeqa
\vskip -0.2in
\beqa
& &S(\rho_{A\!B\!C})+S(\rho_{A\!B\!D})+S(\rho_{A\!C\!D})+S(\rho_{B\!C\!D})+S(\rho_{A})+S(\rho_{B})+S(\rho_{C})+S(\rho_D)\nonumber\\
& &\quad\leq \frac{4}{3}\left[S(\rho_{A\!B})+S(\rho_{A\!C})+S(\rho_{A\!D})+S(\rho_{B\!C})+S(\rho_{B\!D})+S(\rho_{C\!D})\right]\label{S47}
\eeqa

Note the fact that 
\beq
S(\rho_X)\geq E_{GF}(\rho_{X\!Y}),\quad S(\rho_Y)\geq E_{GF}(\rho_{X\!Y})
\eeq
We immediately obtain the bounds on generalized entanglement of formation for tri-party systems in the pure state
\beqa
E_{GF}(\rho_{A\!B\!C}^{\rm P})&\leq&\frac{1}{6}[S(\rho_{A\!B})+S(\rho_{A\!C})+S(\rho_{B\!C})]+\frac{1}{3}[S(\rho_{A})+S(\rho_{B})+S(\rho_C)]\label{3Pupbound}\\
E_{GF}(\rho_{A\!B\!C}^{\rm P})&\geq& \frac{1}{6}[E_F(\rho_{A\!B})+E_F(\rho_{A\!C})+E_F(\rho_{B\!C})]+\frac{1}{3}[S(\rho_{A})+S(\rho_{B})+S(\rho_C)]\label{3Plowbound}
\eeqa
where we have used our definition of the generalized entanglement of formation for any pure tri-particle state (density matrix) $\rho_{A\!B\!C}^{\rm P}$ \cite{My1}
\beqa
{E}_{GF}(\rho_{A\!B\!C}^{\rm P})=\frac{1}{6}\left[E_F(\rho_{A\!B})+E_F(\rho_{A\!C})+E_F(\rho_{B\!C})
+S(\rho_{A\!B})+S(\rho_{A\!C})+S(\rho_{B\!C})+S(\rho_A)+S(\rho_B)+S(\rho_C)\right] \label{GFfor3PP}
\eeqa
In order to compare with the results of the relative entanglement entropy \cite{Plenio2}
\beq
\frac{2}{3}[S(\rho_{A})+S(\rho_{B})+S(\rho_{C})]\geq
E_{R}(\rho_{A\!B\!C}^{\rm P})\geq \frac{1}{3}[E_R(\rho_{A\!B})+E_R(\rho_{A\!C})+E_R(\rho_{B\!C})]+\frac{1}{3}[S(\rho_{A\!B})+S(\rho_{A\!C})+S(\rho_{B\!C})]\label{boundforRE}
\eeq
we first write Eqs. (\ref{3Pupbound}), (\ref{3Plowbound}) and (\ref{boundforRE}) in the forms
\beqa
\frac{1}{2}[S(\rho_{A\!B})+S(\rho_{A\!C})+S(\rho_{B\!C})]\geq
E_{GF}(\rho_{A\!B\!C}^{\rm P})\geq \frac{1}{2}[E_F(\rho_{A\!B})+E_F(\rho_{A\!C})+E_F(\rho_{B\!C})]\\
\frac{2}{3}[S(\rho_{A})+S(\rho_{B})+S(\rho_{C})]\geq
E_{R}(\rho_{A\!B\!C}^{\rm P})\geq \frac{2}{3}[E_R(\rho_{A\!B})+E_R(\rho_{A\!C})+E_R(\rho_{B\!C})]
\eeqa
It is worth to emphasizing that the factors in two bounds inequations are different. This is because in our definition of generalized entanglement of formation, we have taken the equal weight average on various entanglements of formation of subsystem and various entropies of subsystems. If we modify a little our definition by   
\beq
\bar{E}_{GF}(\rho_{A\!B\!C}^{\rm P})\!=\!\frac{1}{3}\left[E_F(\rho_{A\!B})+E_F(\rho_{A\!C})+E_F(\rho_{B\!C})\right]+\frac{1}{6}\left[S(\rho_{A\!B})+S(\rho_{A\!C})+S(\rho_{B\!C})+S(\rho_A)+S(\rho_B)+S(\rho_C)\right] \label{GFfor3PP1}
\eeq
This implies that the extend Bell-states $
\ket{\psi^{\rm EB}_{A\!B}}=\ket{\phi^{\rm B}_{A\!B}}\otimes\ket{\chi_C}$, $\ket{\psi^{\rm EB}_{AC\;1}}=(\ket{0}_A\ket{\chi_B}\ket{0}_C\pm\ket{1}_A\ket{\chi_B}\ket{1}_C)/\sqrt{2}$,$\ket{\psi^{\rm EB}_{AC\;2}}=(\ket{0}_A\ket{\chi_B}\ket{1}_C\pm\ket{1}_A\ket{\chi_B}\ket{0}_C)/\sqrt{2}$,$\ket{\psi^{\rm EB}_{B\!C}}=\ket{\chi_A}\otimes\ket{\phi^{\rm B}_{B\!C}}$ are included in the set of the maximally entangled states since their generalized entanglements of formation are 1. Obviously, we have 
\beq
\frac{2}{3}[S(\rho_{A\!B})+S(\rho_{A\!C})+S(\rho_{B\!C})]\geq
\bar{E}_{GF}(\rho_{A\!B\!C})\geq \frac{2}{3}[E_F(\rho_{A\!B})+E_F(\rho_{A\!C})+E_F(\rho_{B\!C})]
\eeq
If it is in this case, we can see that for tri-party systems, the upper bound of generalized entanglement of formation that is an equal weight average of entropies for all bi-partite subsystems are larger than the upper bound of relative entanglement entropy that is an equal weight average of entropies for all one-party subsystems, and the lower bound of generalized entanglement of formation that is an equal weight average of entanglements of formation for all bi-partite subsystems are higher than the lower bound of relative entanglement entropy that is the equal weight average of relative entanglement entropies for all bi-partite subsystems. 

However, if we accept Plenio and Vedral's conjecture \cite{Plenio2}
\beq
E_R(\rho_{A\!B\!C})\leq \frac{1}{2}[S(\rho_{A})+S(\rho_{B})+S(\rho_{C})]
\eeq
it seems to need us to keep our original definition. This leads to EPR pairs not to be included in the set of the maximally entangled states. 

Now let us consider the case for four-party systems in the pure state. In general,  its reduced density matrices for tri-party subsystems are the mixed states. Thus, assuming that there are the pure state decompositions 
\beq
E_{GF}(\rho_{X\!Y\!Z}^{\rm M})=\sum_i p_i^{X\!Y\!Z}E_{GF}(\rho_{X\!Y\!Z}[i])\quad (XYZ=ABC,ABD,ACD,BCD)
\eeq
corresponding to the minimized statistical average of generalized entanglements of formation of all component pure states, we can derive out
\beq
E_{GF}(\rho_{X\!Y\!Z}^{\rm M})\leq \frac{1}{6}\sum_i p_i^{X\!Y\!Z}[S(\rho_{X\!Y}[i])+S(\rho_{X\!Z}[i])+S(\rho_{Y\!Z}[i])]+\frac{1}{3}\sum_i p_i[S(\rho_{X}[i])+S(\rho_{Y}[i])+S(\rho_Z[i])]
\eeq
In terms of concavity of von Neumann entropy, it becomes 
\beq
E_{GF}(\rho_{X\!Y\!Z}^{\rm M})\leq \frac{1}{6}[S(\rho_{X\!Y})+S(\rho_{X\!Z})+S(\rho_{Y\!Z})]+\frac{1}{3}[S(\rho_{X})+S(\rho_{Y})+S(\rho_Z)]
\eeq
This means that the upper bound of generalized entanglement of formation for a tri-party system in the mixed state has the same form as one in the pure state. But the lower bound of generalized entanglement of formation for a tri-party system in the mixed state needs to be written as
\beq
E_{GF}(\rho_{X\!Y\!Z}^{\rm M})\geq \frac{1+\gamma_2^{X\!Y\!Z}}{3}[E_{GF}(\rho_{X\!Y})+E_{GF}(\rho_{X\!Z})+E_{GF}(\rho_{Y\!Z})]
\eeq
Here, for simplification, we have denoted the following relation
\beq
\frac{1}{6}\sum_i p_i^{X\!Y\!Z}\left[E_{GF}(\rho_{X\!Y}[i])+E_{GF}(\rho_{X\!Z}[i])+E_{GF}(\rho_{Y\!Z}[i])\right]=\frac{\gamma_2^{X\!Y\!Z}}{3}[E_{GF}(\rho_{X\!Y})+E_{GF}(\rho_{X\!Z})+E_{GF}(\rho_{Y\!Z})]
\eeq
Furthermore, we have
\beqa
& &E_{GF}(\rho_{A\!B\!C})+E_{GF}(\rho_{A\!B\!D})+E_{GF}(\rho_{A\!C\!D})+E_{GF}(\rho_{B\!C\!D})\nonumber\\
& &\quad\leq \frac{1}{3}\left[S(\rho_{A\!B})+S(\rho_{A\!C})+S(\rho_{A\!D})+S(\rho_{B\!C})+S(\rho_{B\!D})+S(\rho_{C\!D})\right]\nonumber\\
& &\quad +S(\rho_{A})+S(\rho_{B})+S(\rho_{C})+S(\rho_{D})\nonumber\\
& &\quad \leq S(\rho_{A\!B})+S(\rho_{A\!C})+S(\rho_{A\!D})+S(\rho_{B\!C})+S(\rho_{B\!D})+S(\rho_{C\!D})
\eeqa
\beqa
& &E_{GF}(\rho_{A\!B\!C})+E_{GF}(\rho_{A\!B\!D})+E_{GF}(\rho_{A\!C\!D})+E_{GF}(\rho_{B\!C\!D})\nonumber\\
& &\quad \geq \frac{2(1+\gamma_2)}{3}\left[E_{GF}(\rho_{A\!B})+E_{GF}(\rho_{A\!C})+E_{GF}(\rho_{A\!D})+E_{GF}(\rho_{B\!C})+E_{GF}(\rho_{B\!D})+E_{GF}(\rho_{C\!D})\right]
\eeqa
where
\beqa
& &2\gamma_2\left[E_{GF}(\rho_{A\!B})+E_{GF}(\rho_{A\!C})+E_{GF}(\rho_{A\!D})+E_{GF}(\rho_{B\!C})+E_{GF}(\rho_{B\!D})+E_{GF}(\rho_{C\!D})\right]\nonumber\\
& &\quad = \sum_{X\!Y\!Z} \gamma_2^{X\!Y\!Z} [E_{GF}(\rho_{X\!Y})+E_{GF}(\rho_{X\!Z})+E_{GF}(\rho_{Y\!Z})]
\eeqa
Similarly,
\beqa
& &E_{GF}(\rho_{A\!B})+E_{GF}(\rho_{A\!C})+E_{GF}(\rho_{A\!D})+E_{GF}(\rho_{B\!C})+E_{GF}(\rho_{B\!D})+E_{GF}(\rho_{C\!D})\nonumber\\
& &\quad\leq \frac{3}{2}\left[S(\rho_{A})+S(\rho_{B})+S(\rho_{C})+S(\rho_{D})\right]\nonumber\\
& &\quad\leq S(\rho_{A\!B})+S(\rho_{A\!C})+S(\rho_{A\!D})+S(\rho_{B\!C})+S(\rho_{B\!D})+S(\rho_{C\!D})
\eeqa
Therefore, the upper bound of generalized entanglement of formation for four-party systems in the pure state is just
\beqa
E_{GF}(\rho_{ABCD}^{\rm P})\leq & & \frac{1}{14}\left[S(\rho_{A\!B\!C})+S(\rho_{A\!B\!D})+S(\rho_{A\!C\!D})+S(\rho_{B\!C\!D})\right]\nonumber\\
& &+\frac{2}{21}\left[S(\rho_{A\!B})+S(\rho_{A\!C})+S(\rho_{A\!D})+S(\rho_{B\!C})+S(\rho_{B\!D})+S(\rho_{C\!D})\right]\nonumber\\
& & + \frac{1}{4}\left[S(\rho_{A})+S(\rho_{B})+S(\rho_{C})+S(\rho_{D})\right]
\eeqa
From Eqs. (\ref{S45}-\ref{S47}), it can be written as
\beqa
E_{GF}(\rho_{ABCD}^{\rm P})&\leq 
 &\frac{4}{21}\left[S(\rho_{A\!B})+S(\rho_{A\!C})+S(\rho_{A\!D})+S(\rho_{B\!C})+S(\rho_{B\!D})+S(\rho_{C\!D})\right]\nonumber\\
& & + \frac{5}{28}\left[S(\rho_{A})+S(\rho_{B})+S(\rho_{C})+S(\rho_{D})\right]\nonumber\\
&\leq& \frac{13}{42}\left[S(\rho_{A\!B})+S(\rho_{A\!C})+S(\rho_{A\!D})+S(\rho_{B\!C})+S(\rho_{B\!D})+S(\rho_{C\!D})\right]
\eeqa
The lower bound of generalized entanglement of formation for a four-party system in a pure state reads 
\beqa
E_{GF}(\rho_{ABCD}^{\rm P})&\geq & \frac{5+2\gamma_2}{42}\left[E_{GF}(\rho_{A\!B})+E_{GF}(\rho_{A\!C})+E_{GF}(\rho_{A\!D})+E_{GF}(\rho_{B\!C})+E_{GF}(\rho_{B\!D})+E_{GF}(\rho_{C\!D})\right]\nonumber\\
& &+\frac{1}{6}\left[S(\rho_{A\!B})+S(\rho_{A\!C})+S(\rho_{A\!D})+S(\rho_{B\!C})+S(\rho_{B\!D})+S(\rho_{C\!D})\right]\nonumber\\
&\geq &\frac{5+2\gamma_2}{42}\left[E_{GF}(\rho_{A\!B})+E_{GF}(\rho_{A\!C})+E_{GF}(\rho_{A\!D})+E_{GF}(\rho_{B\!C})+E_{GF}(\rho_{B\!D})+E_{GF}(\rho_{C\!D})\right]\nonumber\\
& &+ \frac{1}{4}\left[S(\rho_{A})+S(\rho_{B})+S(\rho_{C})+S(\rho_{D})\right]\nonumber\\
&\geq& \left(\frac{2}{7}+\frac{\gamma_2}{21}\right)\left[E_{GF}(\rho_{A\!B})+E_{GF}(\rho_{A\!C})+E_{GF}(\rho_{A\!D})+E_{GF}(\rho_{B\!C})+E_{GF}(\rho_{B\!D})+E_{GF}(\rho_{C\!D})\right]
\eeqa
It is easy to obtain the results of the mixed state for a four-party systems. That is
\beqa
& &\frac{13}{42}\left[S(\rho_{A\!B})+S(\rho_{A\!C})+S(\rho_{A\!D})+S(\rho_{B\!C})+S(\rho_{B\!D})+S(\rho_{C\!D})\right]\geq E(\rho_{ABCD}^{\rm M})\nonumber\\
& &\quad \geq \frac{1+\gamma_3(1+\gamma_2)+\delta_2}{6} \left[E_{GF}(\rho_{A\!B})+E_{GF}(\rho_{A\!C})+E_{GF}(\rho_{A\!D})+E_{GF}(\rho_{B\!C})+E_{GF}(\rho_{B\!D})+E_{GF}(\rho_{C\!D})\right]
\eeqa
where assume that the pure state decomposition corresponding to the minimized statistical average of generalized entanglements of formation of all component pure states is taken and define
\beqa
& &\frac{1}{14}\sum_i p_i^{ABCD}\left[E_{GF}(\rho_{A\!B}[i])+E_{GF}(\rho_{A\!C}[i])+E_{GF}(\rho_{A\!D}[i])+E_{GF}(\rho_{B\!C}[i])+E_{GF}(\rho_{B\!D}[i])+E_{GF}(\rho_{C\!D}[i])\right]\nonumber\\
& &\quad =\frac{\delta_2}{6}\left[E_{GF}(\rho_{A\!B})+E_{GF}(\rho_{A\!C})+E_{GF}(\rho_{A\!D})+E_{GF}(\rho_{B\!C})+E_{GF}(\rho_{B\!D})+E_{GF}(\rho_{C\!D})\right]\eeqa
\beqa
& &\frac{1}{14}\sum_i p_i^{ABCD}\left[E_{GF}(\rho_{A\!B\!C}[i])+E_{GF}(\rho_{A\!B\!D}[i])+E_{GF}(\rho_{A\!C\!D}[i])+E_{GF}(\rho_{B\!C\!D}[i])\right]\nonumber\\
& &\quad =\frac{3\gamma_3}{12}\left[E_{GF}(\rho_{A\!B\!C})+E_{GF}(\rho_{A\!B\!D})+E_{GF}(\rho_{A\!C\!D})+E_{GF}(\rho_{B\!C\!D})\right]
\eeqa
For simplicity, the terms related to $\gamma_3, \delta_2$ can be omited.

It is clear that by use of Eqs. (\ref{S31}) and (\ref{S32}) to the multi-party systems, we can find the bounds on generalized entanglement of formation for arbitrary multi-party systems in principle. Of course, it is interesting if the definition of generalized entanglement of formation is an equal weight average of the generalized entanglements of formation and von Neumann entropies taken over all of subsystems. For example, for tri-party system, we point out another possibility. As stated above, this problem depends on our understanding to the set of maximally entangled states. This is still an open question. Even if we takes a unequal weight average of the generalized entanglements of formation and von Neumann entropies taken over all of subsystems, above method can be directly used, but in the bound expressions the coefficients before the entanglements of formation and entropies for bi-partite system will be changed.

\end{document}